\documentclass{desyproc}
\usepackage{float}
\begin{document}
\title{Background model of NaI(Tl) detectors for the ANAIS Dark Matter Project}
\author{{\slshape J.~Amar\'{e}$^{1,2}$, S.~Cebri\'{a}n$^{1,2}$, C.~Cuesta$^{1,2}$\footnote{Present address: Department of Physics, Center for Experimental Nuclear Physics and Astrophysics, University of Washington, Seattle, WA, USA}, E.~Garc\'{i}a$^{1,2}$, M.~Mart\'{i}nez$^{1,2}$\footnote{Present address: Universit\`{a} di Roma La Sapienza, Piazzale Aldo Moro 5, 00185 Roma, Italy}, M.A.~Oliv\'{a}n$^{1,2}$,
Y.~Ortigoza$^{1,2}$, A.~Ortiz de Sol\'{o}rzano$^{1,2}$, C.~Pobes$^{1,2}$\footnote{Present address: Instituto de Ciencia de Materiales de Arag\'{o}n, Universidad de Zaragoza-CSIC, Zaragoza, Spain}, J.~Puimed\'{o}n$^{1,2}$, M.L.~Sarsa$^{1,2}$, J.A.~Villar$^{1,2}$, and P.~Villar$^{1,2}$}\footnote{Corresponding author (pvillar@unizar.es)}\\[1ex] $^1$Laboratorio de F\'{i}sica Nuclear y Astropart\'{i}culas, Universidad de Zaragoza, Calle Pedro Cerbuna 12, 50009 Zaragoza, Spain\\ $^2$Laboratorio Subterr\'{a}neo de Canfranc, Paseo de los Ayerbe s/n, 22880 Canfranc Estaci\'{o}n, Huesca, Spain}

\contribID{familyname\_firstname}

\confID{11832}  
\desyproc{DESY-PROC-2015-02}
\acronym{Patras 2015} 
\doi  

\maketitle

\begin{abstract}
A thorough understanding of the background sources is mandatory in any experiment searching for rare events. The ANAIS (Annual Modulation with NaI(Tl) Scintillators) experiment aims at the confirmation of the DAMA/LIBRA signal at the Canfranc Underground Laboratory (LSC). Two NaI(Tl) crystals of 12.5 kg each produced by Alpha Spectra have been taking data since December 2012. The complete background model of these detectors and more precisely in the region of interest will be described. Preliminary background analysis of a new 12.5 kg crystal received at Canfranc in March 2015 will be presented too. Finally, the power of anticoincidence rejection in the region of interest has been analyzed in a 4$\times$5 12.5 kg detector matrix.
\end{abstract}

\section{The ANAIS experiment and background sources}

The ANAIS project is intended to search for dark matter annual modulation with ultrapure NaI(Tl) scintillators at LSC in Spain, in order to provide a model-independent confirmation of the signal reported by the DAMA/LIBRA collaboration~\cite{dama} using the same target and technique. Two prototypes of 12.5\,kg mass each (referred as D0 and D1), made by Alpha Spectra, Inc. Colorado with ultrapure NaI powder, were taking data at LSC since December 2012 (ANAIS-25 set-up) and a new 12.5 kg module (referred as D2) also built by Alpha Spectra using improved protocols for detector production was added in March 2015 (ANAIS-37 set-up). The goal was the assessment of background and general performance of these detectors. Further description of the ANAIS experiment and these prototypes is given in [2].

The background model of the ANAIS-25 modules has been developed following the same procedure reported in [3]. External background sources from PMTs, copper encapsulation, quartz windows, silicone pads and archaeological lead have been quantified directly by HPGe spectrometry at LSC; also contribution from radon of the inner air volume of the shielding has been considered in the model. Internal contaminations in the NaI(Tl) crystals have been determined from ANAIS-25 and ANAIS-37 data [2] being $^{40}$K (1.25 mBq/kg in all the modules) and $^{210}$Pb (3.15 mBq/kg in D0/D1 and 0.58 mBq/kg in D2) the most relevant contributions in the region of interest. Also $^{129}$I, as for DAMA/LIBRA crystals, has been included in the model. Cosmogenic contributions in the NaI(Tl) crystals have been quantified specifically for ANAIS-25 detectors in [4] and properly considered, being relevant in the long term that of $^{22}$Na. The contribution of these background sources has been assessed by Monte Carlo simulation using the Geant4 code and results are presented in next sections.

\section{ANAIS-25 detectors and the new ANAIS-37 module}

 A detailed description of the ANAIS-25 set-up  including detectors, PMTs and shielding was included in the simulation and spectra at different conditions have been obtained for the different background components. Figure 1 compares the energy spectra summing all the simulated contributions described above with the measured data for ANAIS-25 detectors, considering anticoincidence data. A good agreement is obtained at high energy, but in the very low energy region some contribution seems to be missing. It was found that the inclusion in the model of an additional activity of $\sim$~0.2 mBq/kg of $^{3}$H in the NaI crystals significantly improves the agreement with data at low energy (see figure 2, left). This value is about twice the upper limit set for DAMA/LIBRA crystals, but lower than the saturation activity which can be deduced from the production rate at sea level of $^{3}$H in NaI [5]. Figure 2, right summarizes the different contributions in the region from 1 to 10 keV according to the ANAIS-25 background model.

\begin{figure}[H]
 \centerline{\includegraphics[width=0.47\textwidth]{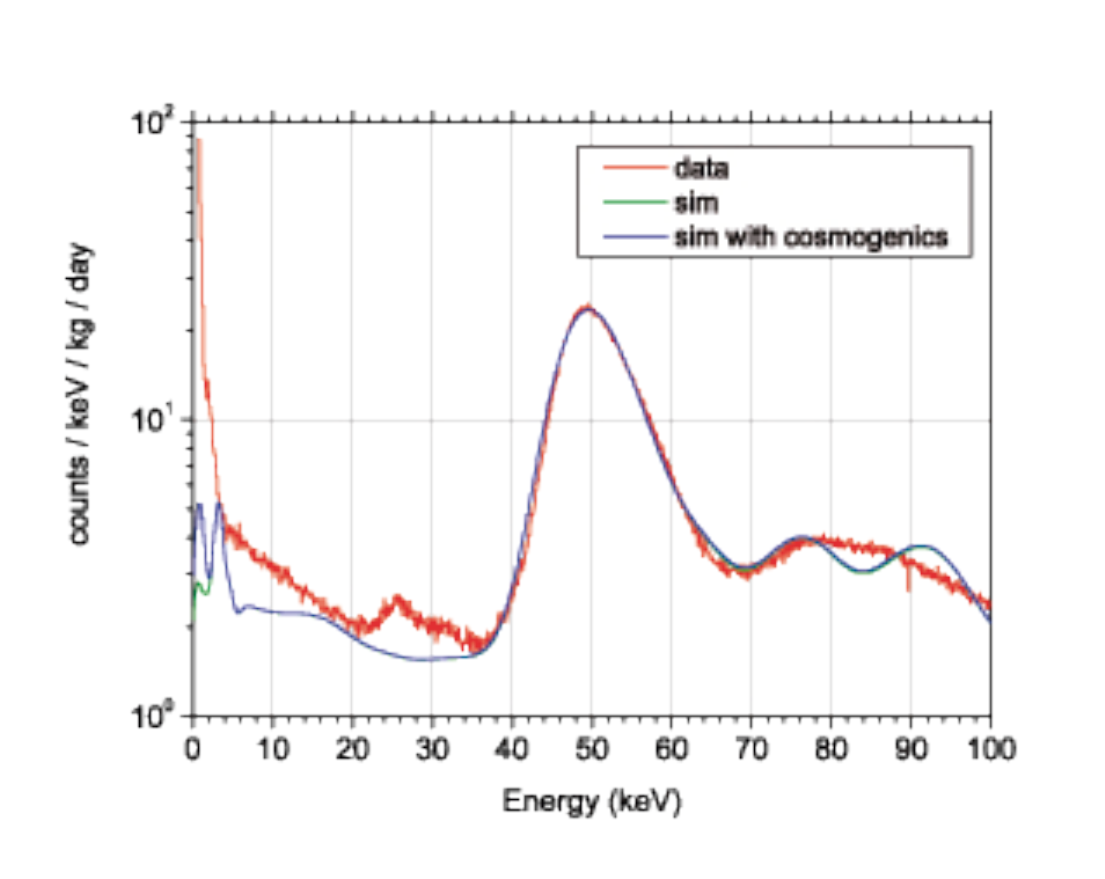}
 \includegraphics[width=0.47\textwidth]{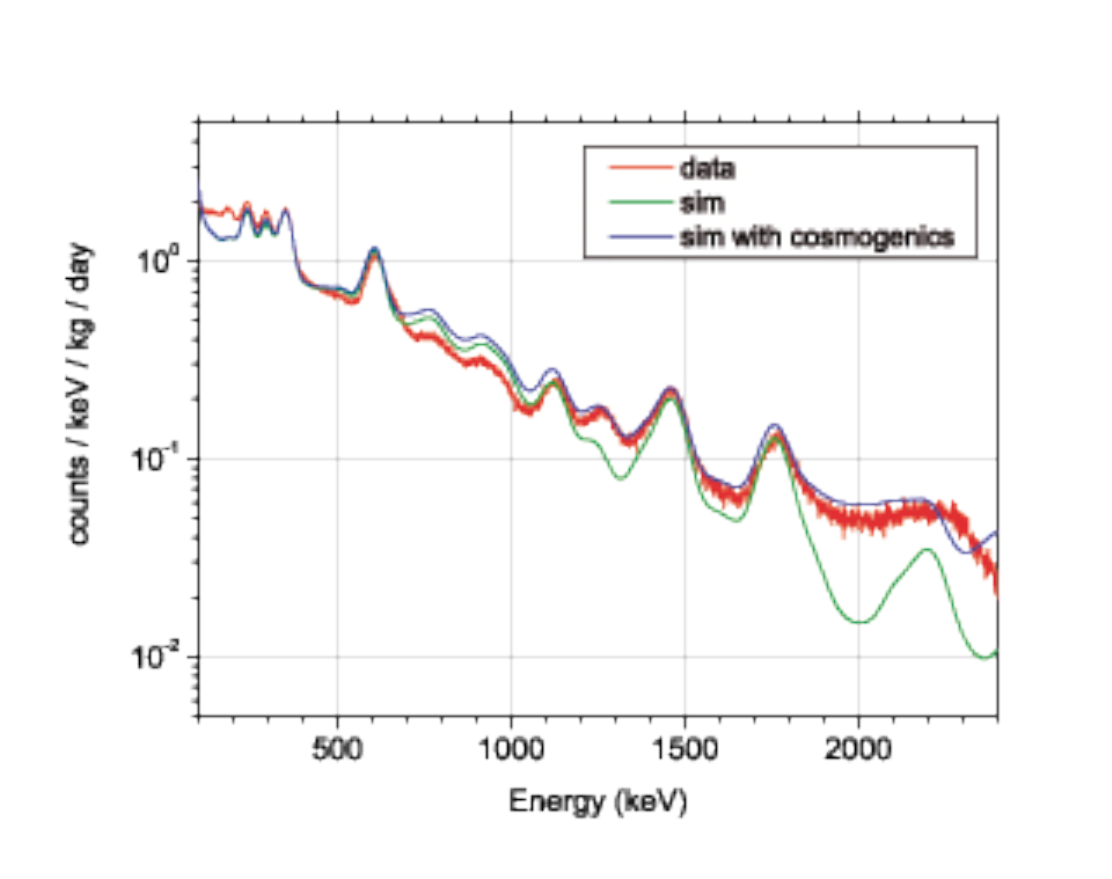}}
  \caption{Comparison of the energy spectra summing all the simulated contributions (before and after adding the cosmogenics) with the measured data for ANAIS-25 considering anticoincidence data at low energy (left) and high energy (right).}
   \label{fig:A25_comp}
\end{figure}

\begin{figure}[H]
 \centerline{\includegraphics[width=0.47\textwidth]{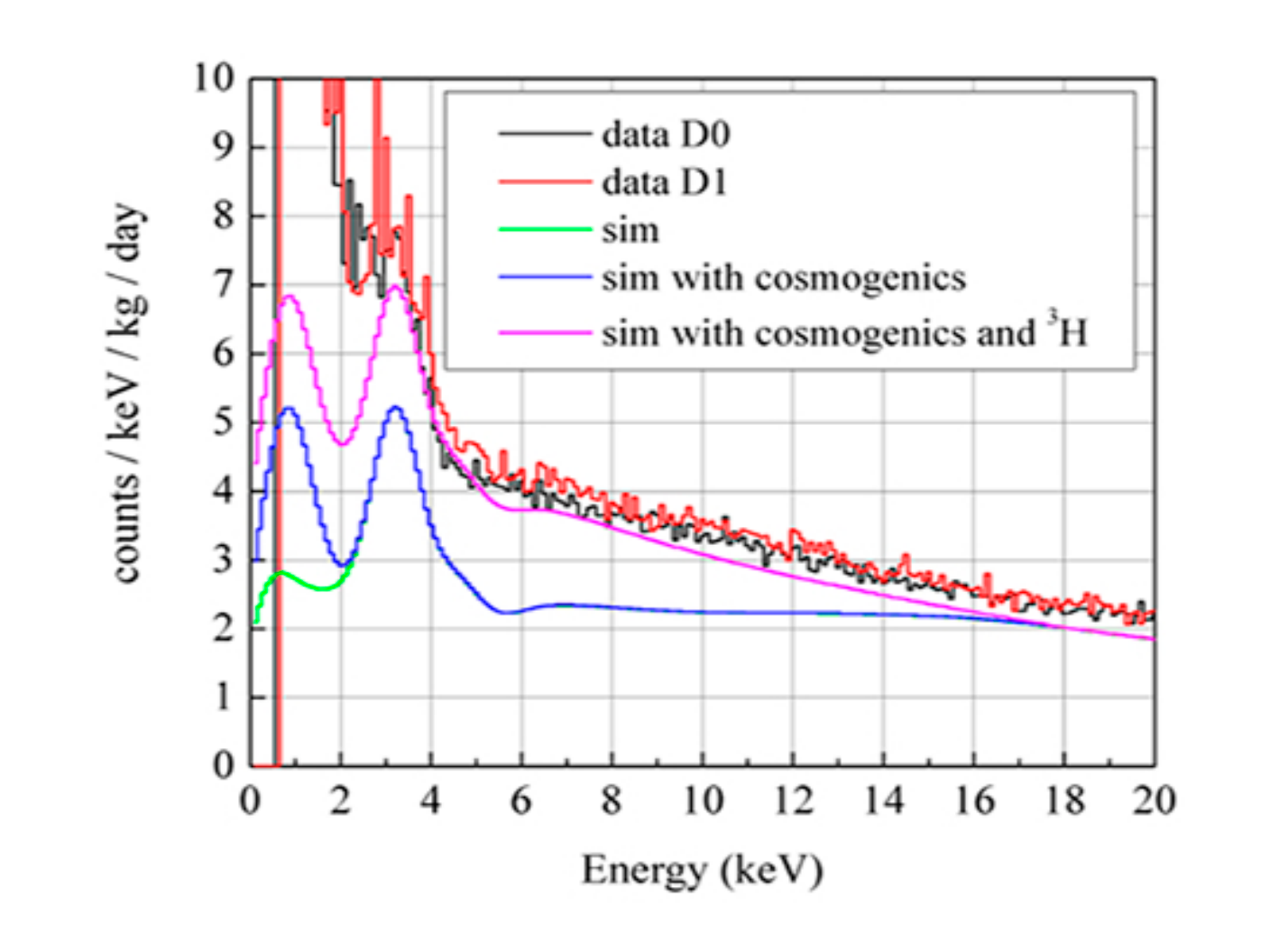}
 \includegraphics[width=0.53\textwidth]{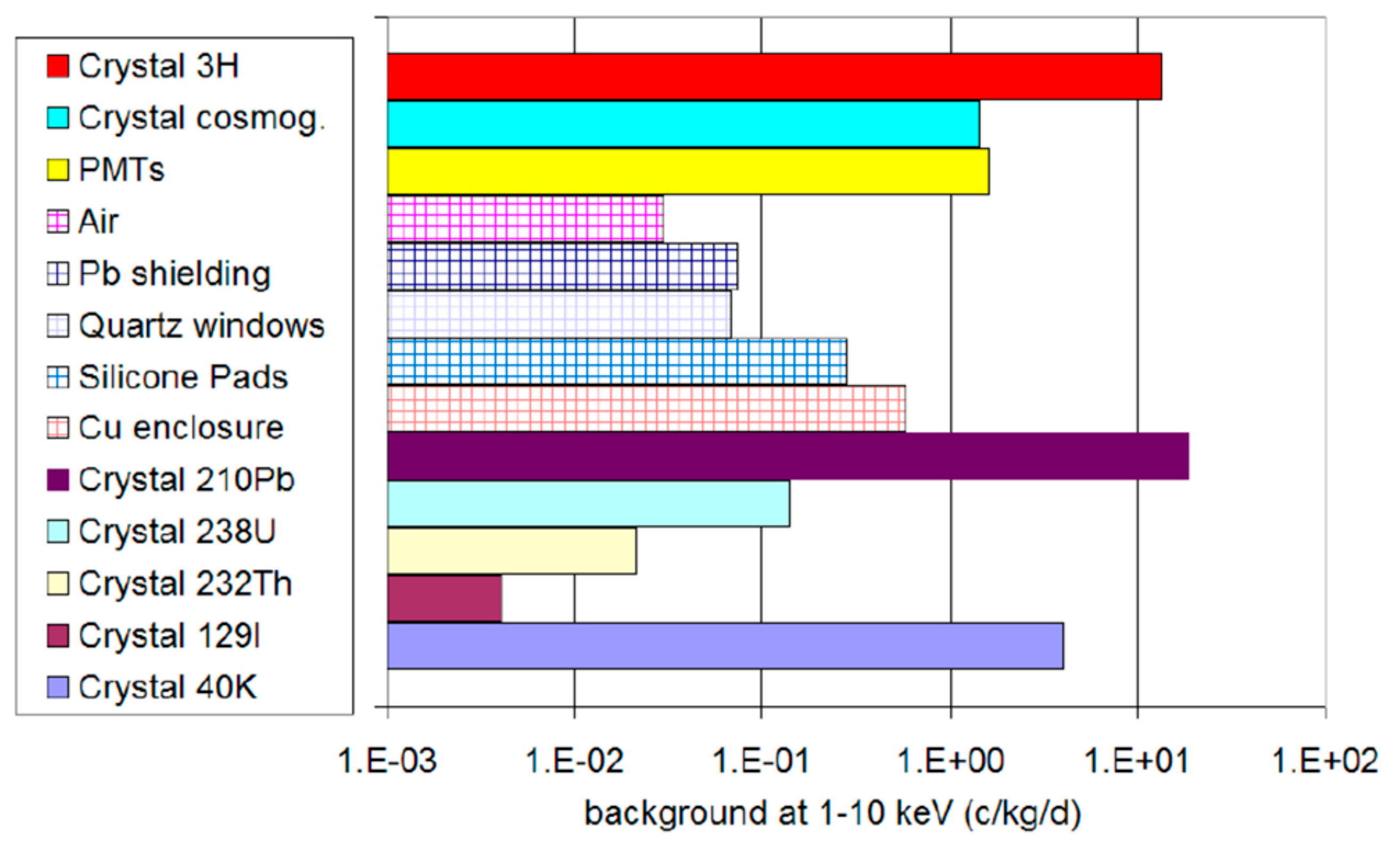}}
  \caption{Effect of the inclusion of $^{3}$H contribution in the very low energy spectrum (left) and different contributions in the region of 1-10 keV according to the ANAIS-25 background model (right).}
   \label{fig:A25_tritio}
\end{figure}

\begin{figure}[H]
 \centerline{\includegraphics[width=0.47\textwidth]{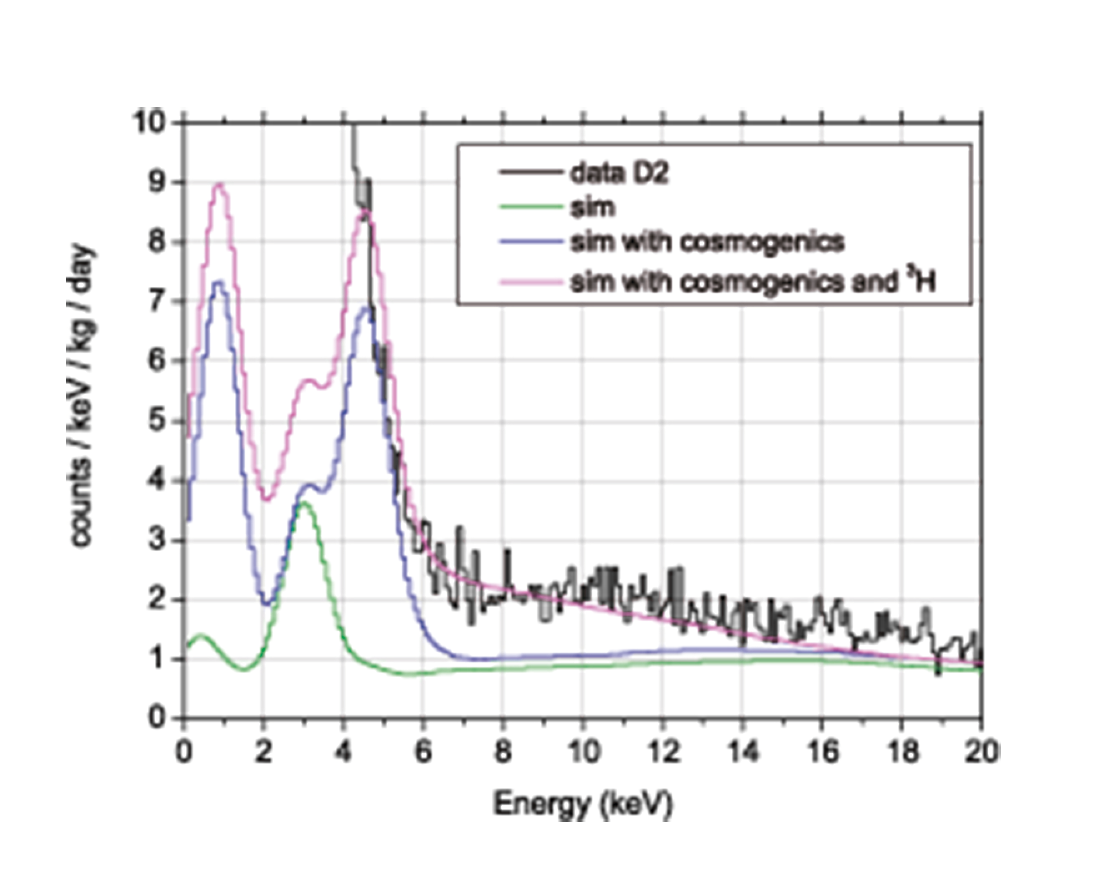}
 \includegraphics[width=0.50\textwidth]{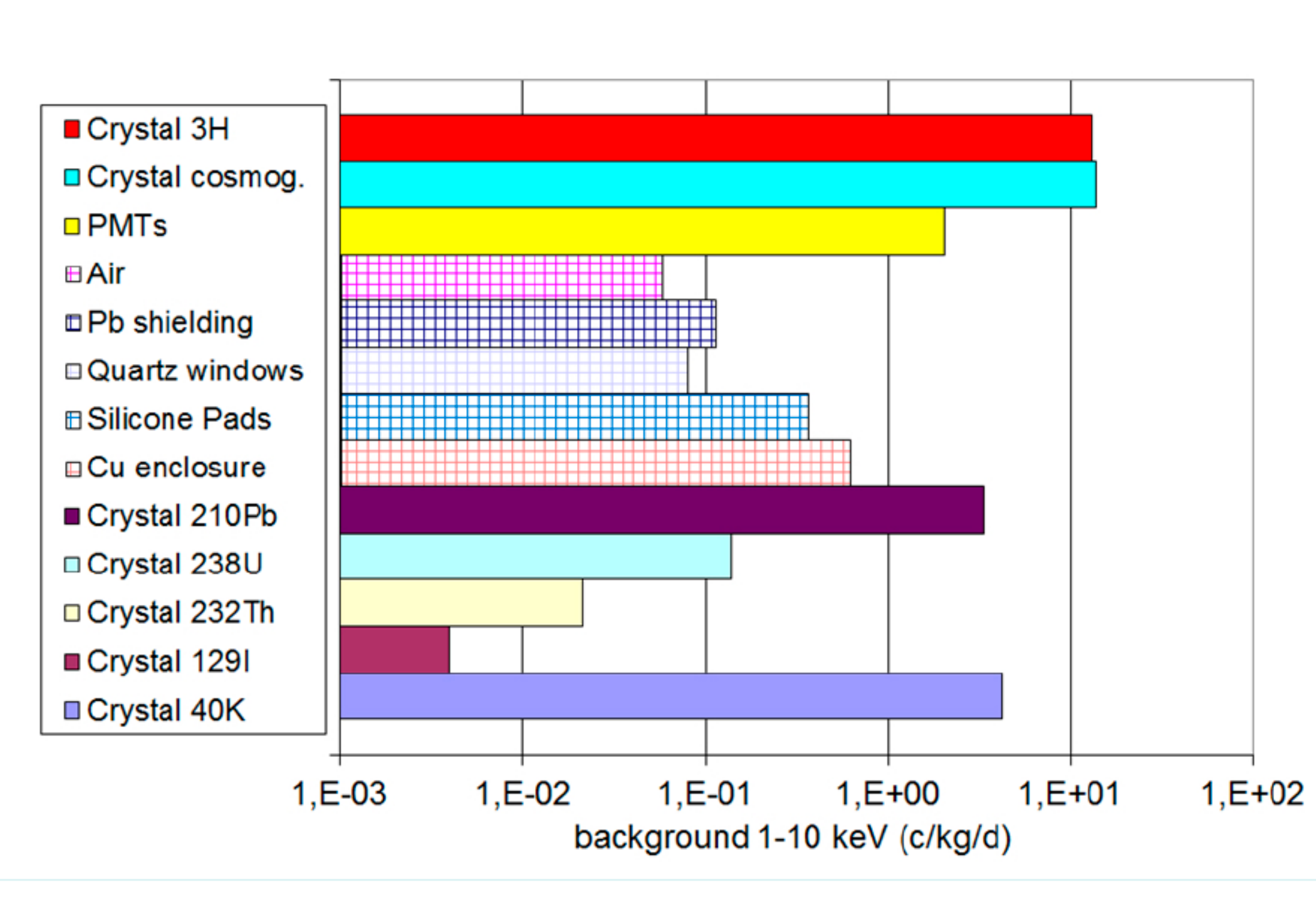}}
  \caption{As figure 2, but for the new module D2 in ANAIS-37 set-up (considering ANAIS-25 activity). No event selection protocols (as those reported in [6]) have been applied to D2 data yet.}
  \label{fig:A37foto}
\end{figure}

The new Alpha Spectra module grown under improved conditions in order to prevent radon contamination was mounted together with  the previous ones forming the ANAIS-37 set-up. The new module (D2) is placed in between the two ANAIS-25 modules (D0 and D1) to maximize the coincidence efficiency for the potassium determination. Very preliminary results are presented here according to the first 50 days of live-time. A total alpha rate of 0.58 $\pm$ 0.01 mBq/kg in the new module D2, determined through pulse shape analysis, is a factor 5 lower than alpha rate in ANAIS-25 modules (3.15 mBq/kg).
Data above 5 keV are well reproduced by our background model (see Figure 3, left) considering $^{210}$Pb activity reduced with respect to D0-D1 in the same factor than alpha rate is reduced and considering the cosmogenic contribution in D2 is still important in the region of interest. Except $^{22}$Na and tritium, these contributions should strongly decay in a few months.

\section{Towards ANAIS}
A good description of the measured background data of ANAIS-25 and ANAIS-37 prototypes has been achieved, being the main contributions in the region of interest the continua from $^{210}$Pb and $^{3}$H and the peaks from $^{40}$K and $^{22}$Na, all coming from the NaI(Tl) crystals. The latter ($^{40}$K and $^{22}$Na) could be strongly reduced by profiting from anticoincidence. Anticoincidence rejection power of different experimental configurations is under study. Just as an example, figure 4 left illustrates  the background reduction expected for the $^{40}$K contribution in the region of interest in a 4$\times$5 detector configuration. A full simulation of the 3$\times$3 matrix of 12.5~kg NaI(Tl) scintillators to be used in the ANAIS experiment is underway, considering also the effect of a liquid scintillator veto (see figure 4, right).

\begin{figure}[H]
 \centerline{\includegraphics[width=0.45\textwidth]{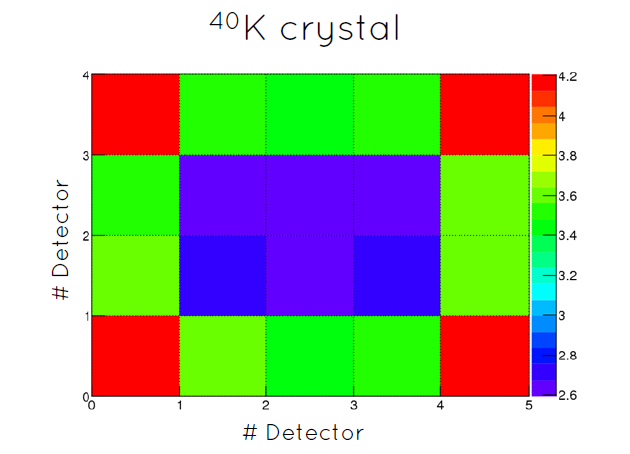}
 \includegraphics[width=0.45\textwidth]{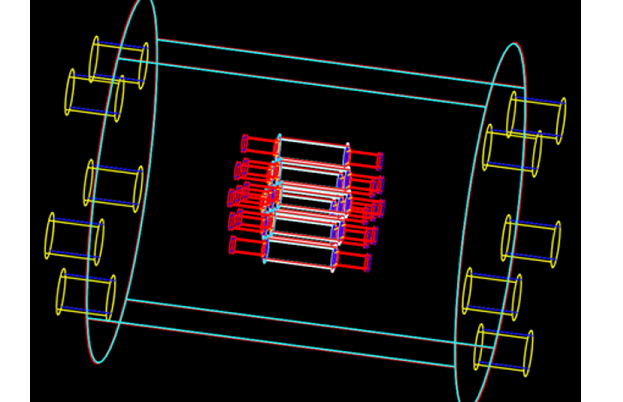}}
  \caption{Distribution of background level below 10 keV (c/kg/d) in anticoincidence at each crystal for $^{40}$K (ANAIS-25 activity) in a 4$\times$5 detector configuration (left) and Geant4 (3$\times$3) matrix of NaI(Tl) scintillators inside a liquid scintillator veto scheme (right).}
  \label{fig:A37foto}
\end{figure}

\section{Acknowledgements}
This work was supported by the Spanish Ministerio de Econom\'{i}a y
Competitividad and the European Regional Development Fund
(MINECO-FEDER) (FPA2011-23749, FPA2014-55986-P), the
Consolider-Ingenio 2010 Programme under grants MULTIDARK
CSD2009-00064 and CPAN CSD2007-00042, and the Gobierno de Arag\'{o}n
(Group in Nuclear and Astroparticle Physics, ARAID Foundation). P.~Villar is
supported by the MINECO Subprograma de Formaci\'{o}n de Personal
Investigador. We also acknowledge LSC and GIFNA staff for their
support.


\begin{footnotesize}

\end{footnotesize}


\end{document}